# P2C2: Peer-to-Peer Car Charging

Prabuddha Chakraborty, Robert C. Parker, Tamzidul Hoque, Jonathan Cruz, and Swarup Bhunia
Department of Electrical & Computer Engineering
University of Florida, Gainesville, FL, USA

*Abstract*— With the rising concerns of fossil fuel depletion and impact of Internal Combustion Engine (ICE) vehicles on our climate, the transportation industry is observing a rapid proliferation of Electric Vehicles (EVs). However, long-distance travel with EV is not possible yet without making multiple halts at EV charging stations. Many remote regions do not have charging stations, and even if they are present, it can take several hours to recharge the battery. Conversely, ICE vehicle fueling stations are much more prevalent, and re-fueling takes a couple of minutes. These facts have deterred many from moving to EVs. Existing solutions to these problems, such as building more charging stations, increasing battery capacity, and road-charging have not been proven efficient so far. In this paper, we propose Peer-to-Peer Car Charging (P2C2), a highly scalable novel technique for charging EVs on the go with minimal cost overhead. We allow EVs to share charge among each other based on the instructions from a cloud-based control system. The control system assigns and guides EVs for charge sharing. We also introduce Mobile Charging Stations (MoCS), which are high battery capacity vehicles that are used to replenish the overall charge in the vehicle networks. We have implemented P2C2 and integrated it with the traffic simulator, SUMO. We observe promising results with up to 65% reduction in the number of EV halts with up to 24.4% reduction in required battery capacity without any extra halts.

## I. Introduction

Electric vehicles have existed for a while, but have never enjoyed mainstream adoption. Now, with the need to reduce our carbon footprint and companies like Tesla, Nissan, and Chevrolet in the picture, the electric vehicle has become more appealing and affordable. Nevertheless, the adoption of EVs remains slow, mainly due to consumer concerns regarding battery life, battery range, and limited access to charging stations [1]. Inefficient charging cycles or complete discharge of a battery reduces its life, making it imprudent to travel the full range provided by the battery without any recharging in the middle [2]. Even though major cities in developed countries have charging stations, the amount is still unable to support a large EV population. Charging stations in remote regions are few and far between. Most of the existing charging stations are Level-2 (220V) which require long waiting periods to charge a vehicle. Level-3 charging stations or DC fast chargers (DCFC) (440V) are a faster alternative; however, they are limited and very expensive to build[3]. With these concerns in mind, researchers have been looking into several potential solutions. Andwari et al. surveyed innovations in EV battery technologies [1], but concluded that the battery range and charging time remains the most critical barrier. Novel solutions like charging via solar-powered roads are not applicable across the geography [4].

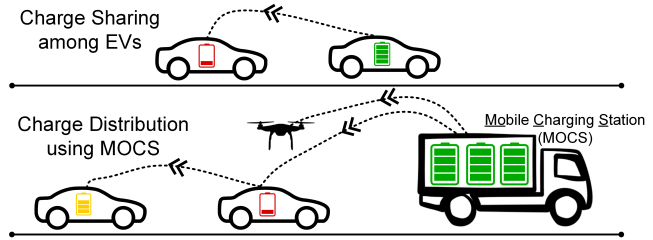

Fig. 1: P2C2 enabled charge sharing among EVs and MoCS-based charge distribution for charging on the go.

In this paper, we propose a scalable peer-to-peer vehicle charging solution that is both low cost and easily to implement with minimal changes to the EVs. As shown in Fig. 1, vehicles will share charge and sustain each other to reach their respective destinations. A set of cloud-based schedulers decides charge providers and receivers. Based on the charge transaction and subsequent reroute decisions, the cloud-based control system instructs the EVs to carry out the charge transfer operations. With this scheme in place, the total charge in the EV network will eventually spread out across all the EVs. However, even in a dynamic network with EVs entering and leaving, we observe through simulation, the total charge of the network will slowly deplete. To keep the EVs in a state of perpetual motion, we introduce Mobile Charging Stations (MoCS), to bring in a considerable amount of outside charge into the EV network. The EVs are then responsible for the fine-grained distribution of the outside charge deposited by the MoCS. We have developed a scheduling algorithm that controls the charge transactions and decides when and where to insert a new MoCS. We quantitatively analyze its effectiveness using SUMO [5] as our traffic simulator. We demonstrate that our algorithm is fast, scalable, and efficient in dealing with battery-related problems present in modern EVs. In particular, we make the following major contributions:

1) We introduce a novel solution to address the electric vehicle charging issue by proposing an on-the-go peer-to-peer charge sharing scheme.
2) We formalize a complete framework to enable electric vehicles for sharing charges guided by a cloud-based control system.
3) We introduce the concept of mobile charging stations, that seamlessly fit into our framework.
4) We propose an algorithm for charge transaction scheduling and MoCS insertion that also controls the EVs for optimal rerouting and charge sharing.
5) We quantitatively analyze the effectiveness of our

solution with an extensive simulation in SUMO [5].

## II. Motivation and Background

### A. Impact of charging issues in EV adoption

*1) Limited range:* The limited battery capacity restricts long-distance driving in EVs. Even with enough charging stations, the travel time is impacted due to frequent, long halts for charging. The price per kilowatt-hour of lithium-ion battery is reducing at a meager rate, making it difficult to increase the battery capacity of EVs without a drastic price increase [1]. High-end EVs such as Tesla Model S and Model X with maximum range of 300 to 370 miles, suffer from high charging times. Even with a 220V charging station, it takes about 10 hours for a full charge [6]. Although 440V stations can reduce the charging time, the amount of charging stations required to support a large EV fleet will be enormous.

*2) Limited stationary charging stations:* The overall number of stationary charging stations are few compared to ICE refueling stations and mostly limited to urban areas. High-end EVs will suffer long charging time from level-1 or level-2 stations. DCFC (Level-3) stations are very few, making it infeasible to sustain a big EV Fleet. Creating large number of DCFC station is financially infeasible as each charging unit costs $10,000-$40,000 [3].

*3) Battery life:* Most of the modern high-end EVs are using Lithium-ion batteries [7]. Complete discharging and charging, or inefficient charging cycles cause the Lithium-ion batteries to age at an accelerated rate [8]. Hence, a long-distance drive without recharging the EV is undesirable for the battery.

### B. Existing solutions to address charging Issues

*1) Better access to fast charging stations:* A brute force solution to the battery range and charging problem is to build a high concentration of very high speed (Level-3) charging stations to allow fast charging anywhere in the world. However, dense and uniformly placed Level-3 stations costing $100,000 each is not feasible. Furthermore, the local power grids must be able to handle the large amount of power that must be transferred in a short amount of time for these stations [9].

*2) Improving battery capacity:* While improving the battery capacity is undoubtedly helpful, it could significantly increase the price of the EV [1]. Besides, it does not solve the core problem of having to stop at a designated station to recharge.

*3) Charging from road:* Charging from the road is an exciting solution to the core problem. However, the solar panels fitted road in Normandy, France produced only 80,000kWh in 2018 and around 40,000kWh by the end of July 2019 due to its inherent dependency over the weather [4]. Converting every road in the world into electric/solar road is a big financial undertaking, rendering the solution infeasible.

## III. Peer-to-Peer Charging Methodology

### A. System overview

To allow efficient charge sharing, we design a cloud-based control system containing a charge transaction scheduling unit, a rerouting unit, and a database for storing the information from EVs. The EVs will interact with each other and the control system, as shown in Fig.2(a). The control system (1) instructs some EVs to share charge with some other EVs, (2) reroutes specific EVs to bring charge providers and receivers together, (3) speed lock EVs to allow seamless charge sharing, (4) detaches a charge provider/receiver for overall network charge optimization. To allow the charge scheduler to operate, the EVs send information to the control system periodically. An example EV-to-EV synchronization for charge sharing is shown in Fig. 2(b) and Fig. 2(c). A big road system can be divided into sections having separate control systems doing the micromanagement. The overall framework will have a global control system that will help in seamless transfer of EVs from one region to another.

Sharing charge between EVs can distribute the total charge in the network among all the entities. But we observe through simulation in Fig. 6(c) that without an outside-the-network charge source, the network will experience a slow overall charge decay increasing the percentage of EV halts. To avoid this problem, we introduce **Mo**bile **C**harging **S**tations (MoCS) which introduces a high volume of charge into the network. Fig.1, shows an MoCS charging a set of EVs in a lane. To identify charge deprived regions, the control unit maintains a charge distribution map that is updated at a regular interval. MoCS are inserted in charge deprived regions if the constraints permit.

### B. Scheduler optimization goals

While computing the reroute, the charge transaction, and the MoCS insertion schedules, the scheduler takes into account certain factors. In our embodiment of the scheduler, we consider the following optimization goals:

1) **Maximize effective charge usage** by analyzing the charge distribution map.
2) **Minimize charging station halts** by sustaining low battery vehicles.
3) **Minimize travel time of all EVs** by limiting the number of rerouting.
4) **Maximize battery life** by taking into account the depth of discharge of each EV.
5) **Prioritize MoCS as provider** over normal EVs.

The final decision of the scheduler is a function of all the optimization parameters, where each parameter can be weighted differently depending on which goals the user wishes to prioritize. Note that the optimization goals we propose are by no means exhaustive and more goals can be added to the scheduler depending on the scenario.

### C. Scheduling Algorithms

The core algorithm for P2C2 scheduling is presented in Algo. 1. The scheduler takes in the charge distribution map ($Charge\_Dist\_Map$) as input and generates a list of instructions ($Instruction\_List$) to be followed by the EVs, MoCS, and MoCS depots. The scheduler acts as an intelligent decision function. We design a method $find\_critical\_evs$ for identifying the EVs in the network with critical battery capacity using the charge distribution map maintained in the cloud control system. In line 3, we use this method to generate the critical EV list ($Crit\_EVs\_List$). The method $find\_prov\_ev$ is designed to identify the best provider EV ($Prov\_EV$) for a given critical EV from all nearby EVs within a user-specified range. This method uses a greedy search algorithm based on a linear weighted function of all the optimization

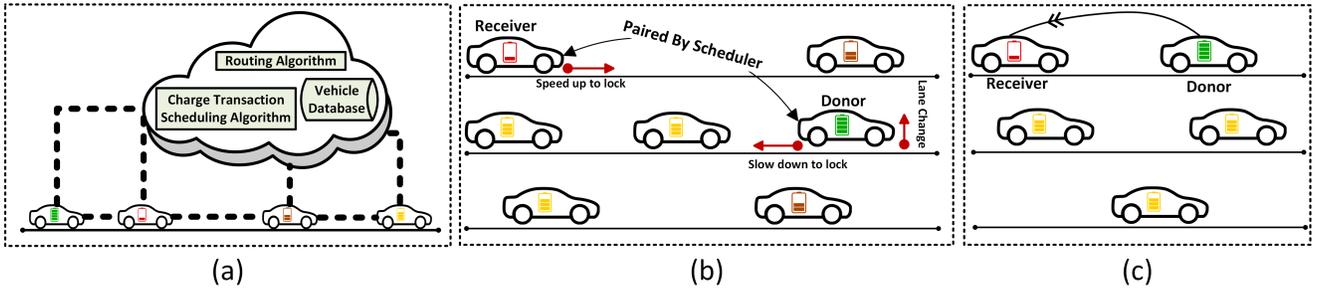

Fig. 2: (a)A system view of P2C2 showing the interaction between the control system and EVs. The control system is located in the cloud facilitates EV-to-EV charge sharing. (b)The paired EVs are being guided by the control system to move closer and come on the same lane. (c)The donor EV is sustaining the EV with critical battery condition.

**Algorithm 1** P2C2 Context-Aware MoCS Scheduler

1: **procedure** GENERATE_SCHEDULE($Charge\_Dist\_Map$)
2:    $Instruction\_List = \emptyset$    ▷ Initialized to empty set.
3:    $Crit\_EVs\_List = find\_critical\_evs(Charge\_Dist\_Map)$
4:    $i = 0$
5:    **while** $i < length(Crit\_EVs\_List)$ **do**
6:      $Prov\_EV = find\_prov\_ev(Crit\_EVs\_List[i])$
7:      $inst = gen\_charge\_tran\_inst(Prov\_EV, Crit\_EVs\_List[i])$
8:      $Instruction\_List.append(inst)$
9:      $i = i + 1$
10:    $Charge\_DR\_List = find\_charge\_dr(Charge\_Dist\_Map)$
11:    $i = 0$
12:    **while** $i < length(Charge\_DR\_List)$ **do**
13:      $ins\_pt = find\_best\_mocs\_ins\_pt(Charge\_DR\_List[i])$
14:      $mocs\_ins\_num = find\_MoCS\_num(Charge\_DR\_List[i])$
15:      $MoCS\_Inst = gen\_mocs\_ins\_inst(ins\_pt, mocs\_ins\_num)$
16:      $Instruction\_List.append(MoCS\_Inst)$
17:      $i = i + 1$
18:    **return** $Instruction\_List$

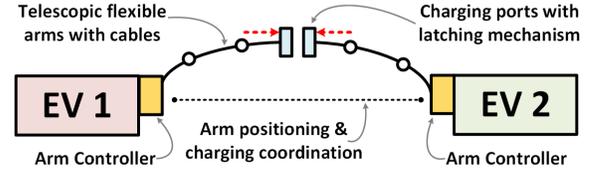

Fig. 3: A physical embodiment of EV-to-EV on-the-go charging mechanism.

goals mentioned earlier. In line 7, we generate the charge transaction instruction ($inst$) required to facilitate the charge transfer. The instruction ($inst$) is appended to the $Instruction\_List$ in line 8. The instructions are targeted towards helping the EVs to come nearby and speed lock. We define a method $find\_charge\_dr$ which finds all the charge deprived regions in the network using linear search. In line 10, the method $find\_charge\_dr$ finds the regions in the road system with a high density of critical EVs. We define two methods $find\_best\_mocs\_ins\_pt$ and $find\_MoCS\_num$ to find out the best MoCS insertion point and amount of MoCS that should be spawned to deal with a particular charge deprived region, respectively. The MoCS insertion point ($ins\_pt$) is selected based on the predicted trajectory of the low battery charge EVs such that the MoCS can easily converge with them. The number of MoCS to be inserted ($mocs\_ins\_num$) is based on the severity (number of critical EVs) of the charge deprived region and the MoCS quota remaining. The function $gen\_mocs\_ins\_inst$ generates the instruction ($MoCS\_Inst$) specifying the amount of MoCS and MoCS insertion location to be sent to the MoCS depot. The complete $Instruction\_List$ is returned in line 18 from the $GENERATE\_SCHEDULE$ method. The instructions generated are sent to the respective MoCS depots and EVs. For the purpose of our simulation, we modify SUMO to emulate MoCS depots and the whole EV network.

### D. Car-to-Car charging mechanism

We envision a safe, insulated, and firm telescopic arm carrying the charging cable. After two EVs lock speed and are in range for charge sharing, they will extend their charging arms, as shown in Fig. 3. The arms heads will contain the charging ports, and they will latch together using either magnetic pads or other means. The arms and the overall charging operation will be coordinated by the respective arm controllers of each EV. This is just one possible realization of the charge transfer mechanism. The entire charging operation can be safely orchestrated if the EVs involved follow a certain predefined protocol. For autonomous/semi-autonomous EVs, the pairing mechanism can be further streamlined. Wireless charging is also possible in the future.

### E. Battery chemistry

Battery to battery charging required for charge sharing is feasible and being actively explored. Products like [10] allow EVs to share charge. Efforts are also being made towards faster-charging batteries. In particular, lithium plating-free charging allows quick recharge at all temperatures without sacrificing the durability of battery cells [11]. Emerging battery technologies such as the aluminum dual-ion battery have been shown to have impressive charging rates and high energy density [12].

## IV. SIMULATION RESULTS PEER-TO-PEER CHARGING

### A. Simulation Setup & Fundamental Observations

To analyze the effectiveness of our cloud control system and the scheduling algorithm, we use an open-source traffic simulator, SUMO (Simulation of Urban Mobility) [5] and integrate the P2C2 scheduler with it. We made modifications to SUMO to support peer-to-peer car charging and MoCS. The P2C2 scheduler communicates with SUMO periodically to gather traffic information and send instructions. We use a 240 km highway to test our method. We run each simulation instance for 5 hours in real-time. We ensure that each EV travels at least 50 km. Each EV weighs 2109 kg with a battery capacity of 75 kWh. Unless otherwise mentioned, the EVs and MoCS enter the simulation with full charge. The weight of each MoCS is 11793 kgs which is the gross vehicle weight rating for a class 6 truck [13]. Each MoCS carries 850 kWh charge and are battery powered themselves. We observe the effect of other parameters such as (1) MoCS-to-EV charge transfer rate, (2) amount of MoCS in the network, and (3) battery capacity reduction of the EVs in later sections.

We test most of our observations on three different traffic scenarios. The internal parameters defining each of these scenarios are as follows:
1) **Light Traffic:** Initially 500 EVs are inserted with a new EV entering the simulation every 4 seconds. A total of 5000 EVs will be inserted over 5 hours.
2) **Medium Traffic:** Initial traffic of 1000 EVs with a new EV entering the simulation every 3 seconds. A total of 7000 EVs will be inserted over 5 hours.
3) **High Traffic:** Initially 2000 EVs are inserted with a new EV entering the simulation every 2 seconds. A total of 11000 EVs will be inserted over 5 hours.

We use a charging rate of 1kW/min for simulation based on a realistic EV-to-EV charging estimate provided in [10]. We consider an EV to be halted when its charge reaches zero. All charge transfer is carried out with 95% efficiency (i.e., 5% loss during transfer).

Fig. 4 illustrates the overall charge distribution in the highway. Each point on the plot indicates the average charge of vehicles in the region. In the charge distribution map shown, we can observe a potential charge deprived region. A few MoCS will probably be inserted in the region depending on the scheduler's decision.

In Fig. 5, we can see the battery charge trend for 6 sampled EVs (red) on the left and 2 sampled MoCS (blue) on the right from the network. The EVs generally experience an initial drop in the battery charge before they are assigned another EV as a provider. After that point, most of the EVs maintain a particular battery level and continue to move perpetually. The purpose of MoCS is to deposit a huge amount of charge in the network quickly; hence, they constantly lose charge as can be seen from the blue plots.

*B. Effect of MoCS Charge Transfer Rate on Halt*

We observe the effect of different MoCS-to-EV charge transfer rate on the percentage of EV halts. 1x charge rate is 1kWh per minute based on [10]. Note that we only change the charge transfer rate between an MoCS and an EV. The EV-to-EV charge transfer rate remains 1kWh per minute throughout the experiment. In Fig. 6(a), we observe that the percentage of halts for all the three traffic scenarios decreases as we increase the MoCS charge transfer rate. If fast charge transfer batteries can be used in the EVs/MoCS, then the effectiveness of P2C2 will be increased. P2C2 charging scheme appears to be more effective in denser traffic scenarios. As can be seen in Fig. 6(a), the percentage of halts for high traffic is least. With more EVs in the network, less amount of rerouting is needed, and an EV with a critical battery state can be quickly assigned to a provider EV which is close by.

*C. Effect of Number of MoCS on Percentage of Halt*

To observe the effect of the number of MoCS in the network on the percentage of EV halts, we set the MoCS-to-EV charge transfer rate to 2x (2kWh per minute), and EV-to-EV charging rate to 1x (1kWh per minute) and vary the limit on the percentage of MoCS in the network. The percentage of MoCS refers to the maximum allowable MoCS for every 100 EVs in the network. In Fig. 6(c), we observe that as we increase the limit of the percentage of MoCS, the percentage of EV halts decreases. So a higher quantity of charge influx also helps in halt reduction.

*D. Charging Time Reduction Analysis*

Based on the battery capacity of the cars used in the simulation, it should take approximately 10 hours to fully charge on the NEMA 14-50 plugs through a 240v outlet [6]. By multiplying the average time charging for each halt with the total number of halts from Table I, we obtain the total charge time for all traffic scenarios. As shown in Table I, the total time spent for stationary charging reduces significantly due to the charge sharing scheme proposed. The % of reduction for P2C2 is calculated compared to the required charge time results for no P2C2 (without). We use MoCS-to-EV charging rate of 2x and 5% MoCS amount limit for obtaining the P2C2 results in Table I.

*E. Battery Capacity Reduction and MoCS Tradeoff*

High capacity batteries in EVs lead to an increase in EV weight and cost. In Fig. 6(b), we observe the effect of reducing the battery capacity of the EVs on the percentage of halts for the medium-traffic scenario. We see the percentage of faults increase as the battery capacity is reduced. There is a trade-off possibility between the amount of MoCS and the battery capacity of EVs. If the amount of MoCS inserted is within 15% of the total EVs in the network, we can reduce the battery capacity of all EVs by 24.4% and still achieve the same amount of halts compared to not using the P2C2 scheme. Therefore, we can reduce the battery capacities of all EVs and have more MoCS running in the system. In future work, we will look into incorporating cost in this trade-off.

V. CONCLUSION

We have presented a novel framework, called P2C2, for charging EVs on the go to address the issue of EV charging infrastructure. P2C2 relies on EV-to-EV coordination as well as a cloud-based guidance system for creating a real-time charge distribution map of the network of EVs and making informed decisions about charge transactions. We incorporate the concept of MoCS - a mobile charging vehicle with a large battery - that can be dispatched to recharge a network of EVs. We have developed a system, algorithm, and method to enable EV-to-EV charging. Using a popular traffic simulator, SUMO, we have realized the P2C2 framework with realistic charging parameters and observed a reduction in the number of halts, and reduction in battery capacity requirements of EVs (thus, leading to reduced cost and weight). Future work will involve augmenting our solution for a heterogeneous network of battery operated entities, like drones and utility robots.

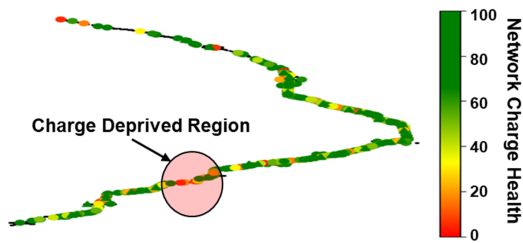

Fig. 4: The charge distribution map maintained by the cloud application at a particular time instance.

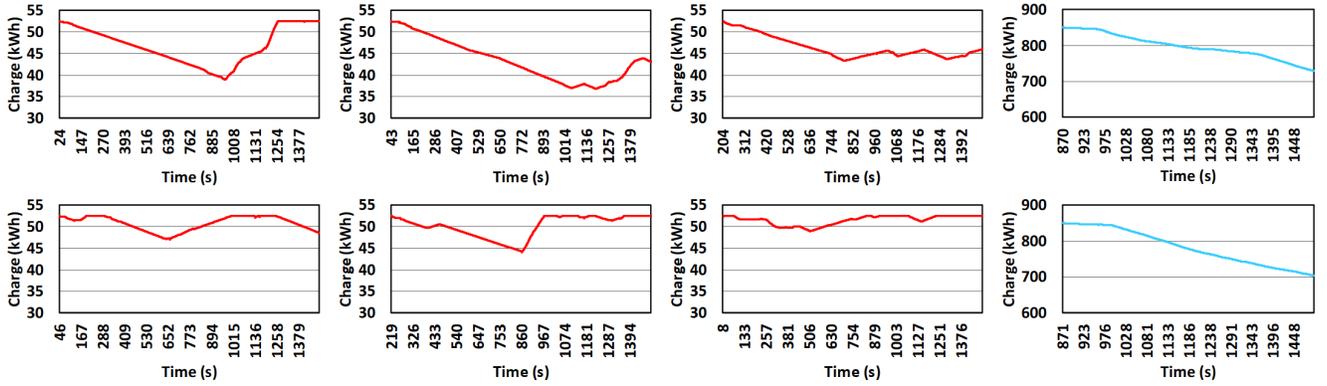

Fig. 5: Change of battery charge level over time for sampled EVs(red) and MoCS(blue) in the network.

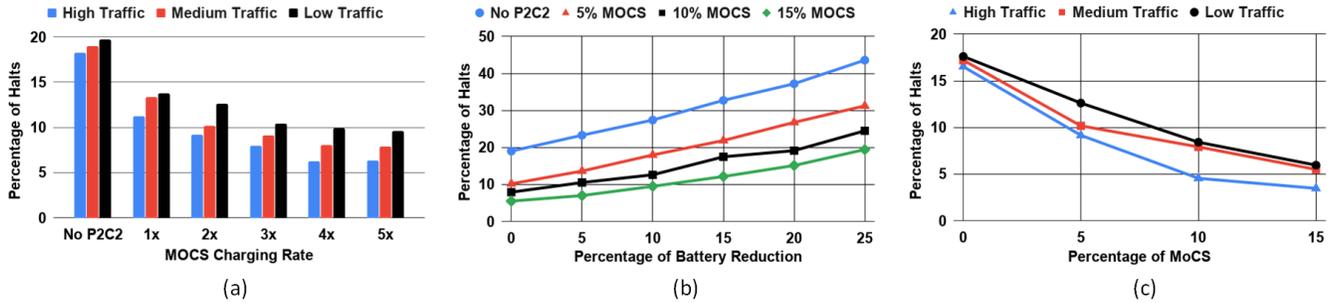

Fig. 6: (a) The percentage of EV halts reduces as the MoCS-to-EV charge transfer rate increases. (b) The percentage of halt increases as we decrease the battery capacity. The halt percentage is less in presence of more percentage of MoCS. (c) The percentage of EV halts reduces as the limit on the percentage of MoCS in the network is increased.

TABLE I: The percentage of halt induced charging time reduction in different traffic scenarios.

|  | Light Traffic | | Medium Traffic | | High Traffic | |
|---|---|---|---|---|---|---|
|  | Baseline | P2C2 | Baseline | P2C2 | Baseline | P2C2 |
| % of Halts | 19.68 | 12.62 | 19.02 | 10.18 | 18.25 | 9.16 |
| Num of EVs | 5000 | 5000 | 7000 | 7000 | 11000 | 11000 |
| Halt Time (hrs.) | 9840 | 6310 | 13314 | 7126 | 20075 | 10076 |
| Halt Time Cut % | - | 35.87 | - | 46.48 | - | 49.81 |